\documentclass[conference]{IEEEtran}
\IEEEoverridecommandlockouts
\usepackage{cite}
\usepackage{amsmath,amssymb,amsfonts}
\usepackage{algorithmic}
\usepackage{graphicx}
\usepackage{textcomp}
\usepackage{color}
\usepackage{xcolor}
\usepackage{tabularx}
\usepackage{listings}
\usepackage{authblk}
\usepackage{hyperref}
\def\BibTeX{{\rm B\kern-.05em{\sc i\kern-.025em b}\kern-.08em
    T\kern-.1667em\lower.7ex\hbox{E}\kern-.125emX}}

\begin{document}

\lstdefinelanguage{DialogueExamples}{
    language = Lisp,
    keywords={},
    otherkeywords={SOPHIE},
    keywords = [2]{User},
    keywordstyle=\color{purple},
    keywordstyle=[2]\color{blue},
}


\title{Validating a virtual human and automated feedback system for training doctor-patient communication skills}

\author{
Kurtis Haut$^*$~~~
Caleb Wohn$^*$~~~
Benjamin Kane$^*$~~~
Tom Carrol$^\dag$~~~
Catherine Guigno$^*$~~~\\
Varun Kumar$^*$~~~
Ron Epstein$^\dag$~~~
Lenhart Schubert$*$~~~
Ehsan Hoque$^*$~~~
\smallskip 
\\
$^*$University of Rochester Department of Computer Science
\\
$^\dag$ University of Rochester Medical Center

}

\maketitle

\begin{abstract}
\footnote{Kurtis Haut and Caleb Wohn share the distinction of first authorship.}Effective communication between a clinician and their patient is critical for delivering healthcare maximizing outcomes. Unfortunately, traditional communication training approaches that use human standardized patients and expert coaches are difficult to scale. Here, we present the development and validation of a scalable, easily accessible, digital tool known as the Standardized Online Patient for Health Interaction Education (SOPHIE) for practicing and receiving feedback on doctor-patient communication skills. SOPHIE was validated by conducting an experiment with 30 participants. We found that participants who underwent SOPHIE performed significantly better than the control in overall communication, aggregate scores, empowering the patient, and showing empathy ($p < 0.05$ in all cases). One day, we hope that SOPHIE will help make communication training resources more accessible by providing a scalable option to supplement existing resources.
\end{abstract}


\section{Introduction}
\label{intro}

\begin{figure*}[t]
    \centering
    \includegraphics{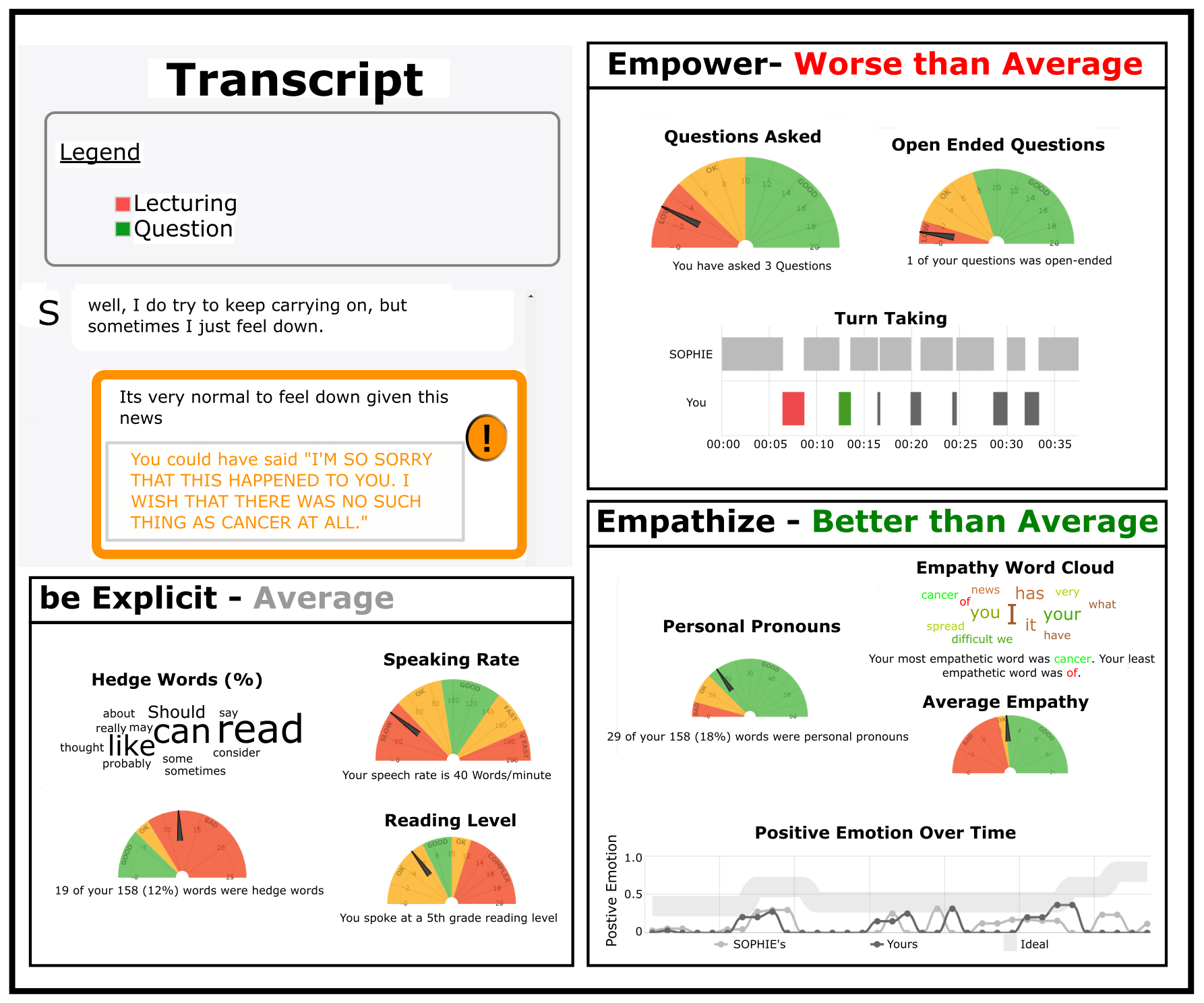}
    \caption{\textbf{SOPHIE Feedback System} - The feedback is divided into four sections; Transcript, Empower, be Explicit, and Empathize. The upper left contains the transcript with embedded conversational suggestions. The Empower section contains the metrics number of questions asked, number of open-ended questions asked, and turn-taking with lecture and question coloring. The be Explicit section contains the metrics hedge words percentage with word cloud, speaking rate and reading level. The Empathize section contains the metrics personal pronouns percentage, average empathy score (1-7) with word cloud, and positive emotion (sentiment) over time graphs for the user, SOPHIE, and the ``ideal'' sentiment trajectory.}
    \label{fig:feedback}
\end{figure*}

 60\% of late-stage cancer patients leave their doctor's office without fully understanding their prognosis \cite{hagerty2005communicating} and 79\% of patients feel emotionally unsupported by their doctors \cite{korsch1972doctor}. Past research has shown that poor communication by doctors leads to lower quality healthcare outcomes at a higher cost \cite{korsch1972doctor}\cite{ha2010doctor}\cite{riedl2017influence}\cite{stewart1995effective}\cite{begum2014doctor}. Unfortunately, low cost communication training videos or reading materials have been shown to have little effect \cite{arnold1994accelerating}\cite{ijaz2017virtual}. Training courses using standardized patients (SPs) are a viable remedy widely used in medical schools  \cite{fiscella2007ratings}\cite{teherani2008can}. For example, our institution offers interdisciplinary workshops for practicing patient care professionals (e.g., physicians, nurses, advanced practice providers, social workers, and chaplains) through the Advanced Communication Training (ACT) program\cite{carroll2021re}, which  teaches the MVP (Medical situation, Values, Plan) paradigm and emphasizes the 3 E skills: Empower, be Explicit, Empathize skills\cite{horowitz2020mvp}. Receiving feedback has been found to improve the communication skills of clinicians. For example, feedback from communication coaching experts based on recorded interactions with real patients has been shown to improve a clinician's ability to empathize with their patient and empower them by eliciting questions \cite{10.1001/jamainternmed.2023.0629}. However, due to the cost and limited availability of human SPs and coaches who can provide relevant feedback, these traditional approaches are hard to scale. The need for a scalable solution is compounded by the diminishing effects of communication training over the course of a physician's career \cite{dimatteo1998role}.

We developed SOPHIE (Standardized Online Patient for Health Interaction Education) \cite{ali2019online} to address this need. SOPHIE is a fully automated web-based system allowing medical professionals to have a conversation with a virtual human using their computer's speakers and microphone. After the conversation, the system automatically analyzes the transcript to provide immediate, quantified, and personalized feedback. 

Using virtual patients for educating health professionals is not a new concept \cite{cook2010computerized}. Prior work has shown the value of virtual patients in practicing empathy in a low stress environment \cite{kleinsmith2015understanding}, and much promise is granted to using virtual patients as a cost-effective pedagogical approach for developing countries \cite{dewhurst2009online}. The recent advancements in avatar generation and natural language understanding have opened up exciting possibilities for creating more realistic, interactive systems capable of providing user feedback that was previously not possible.

Indeed, the feedback component of the SOPHIE system represents a distinct contribution (see Fig. \ref{fig:feedback}). Although prior work has shown that receiving feedback helps clinicians improve their communication skills \cite{custer2019development}, there are few existing tools to generate feedback automatically \cite{CUFFY2020103589}. Our feedback system is unique in that it utilizes the previously validated MVP/3E’s model of doctor-patient communication. It provides a quantitative analysis of the conversation for medical professionals to review, as well as text recommendations for improvement.

We validated the feasibility of this system in a experiment with 30 participants (See Fig. \ref{fig:experiment}). We found that participants who underwent the educational intervention with SOPHIE performed significantly better in overall communication and achieved higher aggregate scores compared to participants who did not ($p < 0.05$). We also observed statistically significant results for empowering the patient and showing empathy. We hope the SOPHIE system will eventually be utilized as a scalable solution to supplement existing communications training or as a low-cost alternative for resource deprived communities.

\section{Methods}
\subsection{The SOPHIE System}
The educational intervention with the SOPHIE system has three components. The user begins by watching an instructional video about the MVP/3E’s communication paradigm followed by viewing a tutorial video on how to use the SOPHIE system. The final component of the intervention is two conversations with SOPHIE, including feedback after each conversation. SOPHIE portrays an older female patient with advanced lung cancer who is seeking information about the prognosis. The feedback page is split into 4 main sections: a transcript, and one section for each of the three E’s (see Fig. \ref{fig:feedback}). The transcript section allows the user to review their conversation. Segments of the conversation where the medical professional engaged in lecturing (i.e. spoke for too long) are given a red background, and segments where the medical professional empowered the patient by asking a question are given a green background. Some  segments in the transcript display suggestions for open-ended questions or empathetic statements that the medical professional could have used. The feedback system was developed through an iterative design process with close collaboration between programmers and palliative care specialists, and many of the metrics are based on statistical analysis of doctor-patient communication, as discussed in \cite{ali2021novel}.

\subsection{Dialogue Management}
\begin{figure*}[t]
    \centering
    \includegraphics{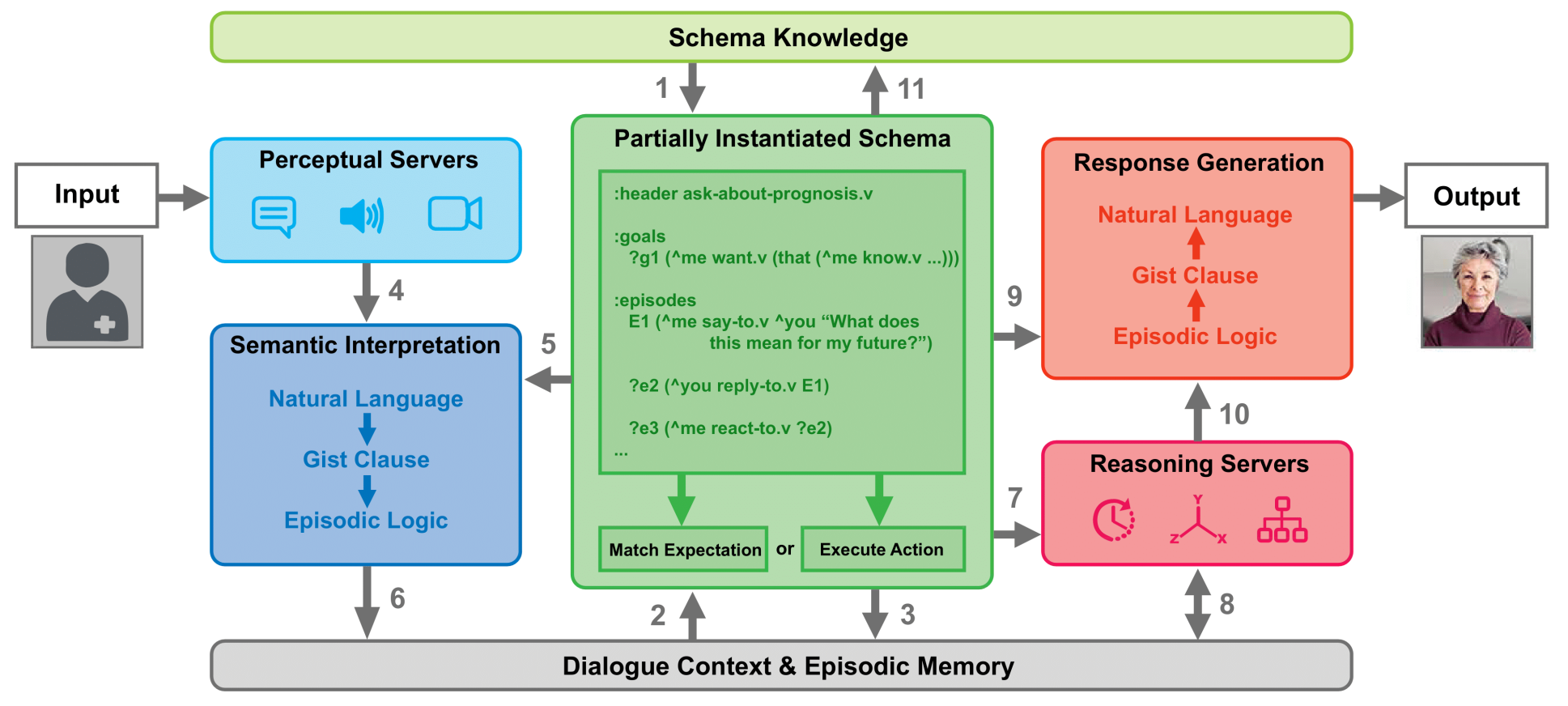}
    \caption{\textbf{Eta Dialogue Management Architecture} - The dialogue manager relies on a database of general schema knowledge including dialogue schemas (top), as well as dialogue context and episodic memory (bottom). The dialogue manager interleaves processing of several submodules for processing input (blue; left), guiding system behavior through dynamic instantiation of a dialogue schema (green; center), and generating output (red; right). The numbered edges represent the flow for how the system interacts with the user.}
    \label{fig:eta}
\end{figure*}

SOPHIE's dialogue manager uses a symbolic, schema-based approach. Although LLMs have recently achieved impressive results \cite{openai2023gpt4}\cite{bubeck2023sparks}, at the time of development, they were deemed ill-suited to this task for a variety of issues.
Bender et al. argued that large language models (e.g., the current state-of-the-art) are generally insufficient for true language understanding as well as carry their own risks and potential ethical issues \cite{bender2020climbing}. Large language models also come with the additional risk of going ``off the rails" of the conversation parameters which makes them unpredictable and difficult to control. Without having the ability to control the dialogue, presenting the user with opportunities to practice specific communication skills poses a real challenge in application consistency. 
As a result of these issues, we chose to take a symbolic approach. The conversations with SOPHIE are driven by eta, which uses flexible, modifiable dialogue schema (i.e., expected event types expressed as conversational statements) to imitate natural human conversations. The dialogue manager dynamically plans and enacts the conversation in real time by combining a user interpretation process with these dialogue schema\cite{razavi-2016-lissa}\cite{razavi-2019-lissa}\cite{razavi-2017-managing-dialogue-schemas}. See figure \ref{fig:transcript} for an example of the dialogue and see Fig. \ref{fig:eta} for an overview of the dialogue manager's architecture). The user interpretation process is handled by a set of pattern transduction rules that map user utterances into simplified context-independent ``gist clauses” given the immediate context of the preceding dialogue turn. The gist clause provides an explicit representation of the meaning of the user’s utterances that the system can then respond to. Response generation, which is also handled by a pattern transduction process, can involve the selection of a particular reaction by the system to the user’s gist clause, or the invocation of a new schema (e.g., the system may invoke a schema for SOPHIE discussing her medical concerns if asked a relevant question by the doctor). In the case where the system fails to extract a gist-clause, it may either ask the user to repeat and clarify their utterance, or give a generic default reaction specific to the current schema. Ultimately, a schema guided approach to dialogue management was chosen over using a large, neural language model (such as GPT3) in order to have more control over the dialogue.

\subsection{Quantifying Complex Human Communication Skills}
 \textbf{Empower -} We quantified empowerment using three key metrics; Questions-asked, open-questions asked and turn-taking. The need to quantify both types of questions was made apparent after consultation with Oncologists and Palliative Care specialists at University of Rochester Medical Center (URMC). A closed question helps the medical professional check patient understanding of the medical situation and prognosis while an open-ended question gives the patient an opportunity to reveal more sensitive information about their emotional state. We used expression matching to determine what type of question was asked and kept track of the total number in each category. Questions are useful for quantifying empowerment because this inevitably invites the user to take control of the conversation and thus tends to empower them. For example, by asking a question, the patient can express their concerns and voluntarily reveal external factors that would otherwise be hidden to the medical professional. 
 
 Turn-taking was quantified by keeping track of the total time for both the user and SOPHIE respectfully. Whether or not the user was lecturing for their turn was based off a previous voice study \cite{ali2021novel} and informally takes place when a medical professional is speaking for too long. This quantifies empowerment because unequal turn taking could signal a lack of empowerment for the user if their turns are too long and frequent in comparison to SOPHIE's.

\textbf{Empathize - } We quantified empathize using three key metrics; sentiment trajectory, empathy word cloud and personal pronouns. Sentiment trajectory was computed based off the work of Ali et al. \cite{ali2021novel}, who analyzed sentiment in the VOICE dataset led by Sen et al. \cite{sen2017modeling}. They defined a ``sentiment trajectory" as an average sentiment vector across time. Sentiment was computed using VADER \cite{hutto2014vader} and is on a -1 to +1 scale. They used k-means clustering to identify ``sytles” of sentiment trajectory and logistic regression to identify if any style was associated with good conversation outcomes where ``good" is determined by the level of patient prognosis understanding. The best style was ``dynamic,” having high sentiment early on in the conversation, low sentiment in the middle (likely to match patient sentiment after hearing the prognosis), and high sentiment at the end (likely to express encouragement, care and support). Our idea was that perhaps some of the complexity of empathy could be captured and quantified overall using this sentiment trajectory.

The empathy word cloud was computed using work from Sedoc et al. \cite{sedoc2019learning} who developed a lexicon which maps 10k words to empathy ratings on a 1-7 scale using a Mixed-Level, Feed Forward Network. We mapped every word spoken by the user through this lexicon and computed the average empathy from that. The word cloud is creating using the 15 most frequent words. We hoped that this would roughly quantify empathy for the entire conversation, but recognize the weakness in its inability to quantify statements on a sentence-level (which is where most empathy is likely to take place).

Personal pronouns were chosen to quantify empathy based of work by Sen et al. in 2017 \cite{sen2017modeling} who used LIWC to analyze the correlation between different categories of words and patient ratings of doctors in the VOICE dataset. Interestingly, they found that higher rankings were correlated with the use of personal pronouns (e.g. I, You, etc.). We computed pronoun usages using NLTK part-of-speech tagging \cite{bird2006nltk}. Intuitively, using pronouns makes a conversation more empathetic by appearing more personable as compared to generic, disease specific speech.

\textbf{be Explicit - } We quantified being Explicit using three metrics; speaking rate, reading grade and hedge words. Speaking rate was found to be associated with patient understanding \cite{ali2021novel} and is related to being explicit (e.g., if a physician speaks too slowly, or too quickly they probably are not communicating in an explicit manner). This is computed simply by taking the average words spoken per minute. 

Reading grade was chosen quantify the complexity of the speech to address the issue of doctors using too much medical jargon when communicating a prognosis. Complex speech makes it more difficult for patients to understand their medical situation and we make the assumption that less complex speech is more explicit. We computed the reading grade using the standard measure of the Flesch-Kincaid readability test\cite{si2001statistical} which outputs the linguistic complexity in terms of U.S. school grades (e.g., 1st grade to 12th grade).

Hedge words were computed using a simple list of hedge words established from prior work by counting the percentage of a user's words that appeared on the list. The most frequent 10 hedge words are made note of and a word cloud is constructed. The 2020 Horowitz et al’s paper on the MVP paradigm which SOPHIE attempts to model mentions avoiding hedging as a key part of being explicit  \cite{horowitz2020mvp}. Thus, measuring hedging helps to quantify this skill directly for the user.

\subsection{Experiment}
\begin{figure*}[t]
    \centering
    \includegraphics{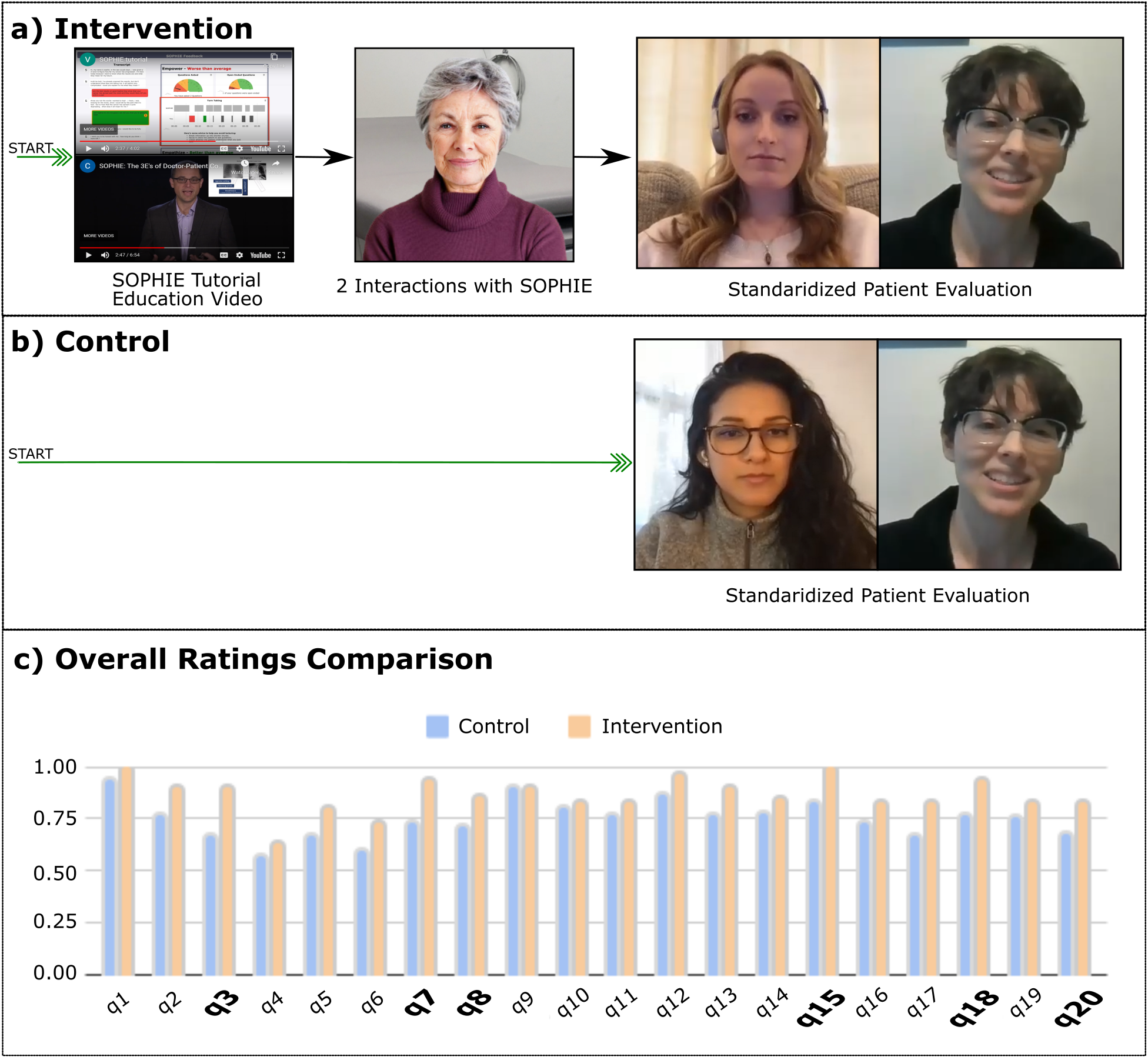}
    \caption{\textbf{Experiment with 30 participants} - a) Intervention group, underwent educational intervention with SOPHIE before speaking to SP. b) Control group, received no training before speaking to SP. c) Overall ratings comparison between control (blue) and intervention (tan) \textbf{bold} denotes significant differences. The numbers have been normalized to a 0-1 scale with 1 being ``good.'' The raw numbers and full question text can be found in Table \ref{tab:results} by looking up the question ID. (Images of participants used with permission).} 
    \label{fig:experiment}
\end{figure*}

We conducted an experiment consisting of 30 participants with medical backgrounds (12 medical students, 9 nurses, 4 internal medicine residents, 2 physician's assistants, 2 psychologists, and 1 hospital chaplain). Participants were randomized (1:1 ratio) into intervention and control groups, stratified by professional/training background. The intervention group underwent the educational intervention with the SOPHIE system while the control group received no training (see Fig. \ref{fig:experiment}a and \ref{fig:experiment}b).

After the educational intervention, we evaluated the communication skills of the participants using human standardized patients (SPs). Every participant had a conversation with a SP via a video call. Immediately after each interaction, the SP rated the conversation using a standard scale developed with assistance from palliative care specialists. These ratings were statistically compared using a Mann Whitney U test to determine whether there were any significant differences between the the two groups as a result of undergoing the educational intervention with the SOPHIE system (see Fig. \ref{fig:experiment}c). Additionally, participants in the treatment group completed a UI/UX survey designed to inform future iterations of the SOPHIE System. 

\subsection{Justification for Evaluation Metrics}
\textbf{SP Rating Scale -} Our rating scale was developed in close collaboration with URMC Oncologists and Palliative Care Specialists. There was no existing rating scale that was appropriate as is for our experiment although prior work exists \cite{zill2014measurement}. We wanted to measure whether a clinician improved based on behaviors the SOPHIE system was designed to give feedback on and reinforce. The full rating scale can be seen in table \ref{tab:results} and is based on behaviors the human SP observes during their interation.

\textbf{UX/UI Rating Scale -} We broke the UX scale down into three components; namely system usability, virtual human and dialogue (see Figure \ref{fig:UX_responses}). We chose a representative sample of system usability statements from the well-established System Usability Scale (SUS)\cite{bangor2008empirical}\cite{lewis2018system}. The statements for the virtual human and dialogue sections were developed by the research team after robust discussions. We wanted to evaluate the realism of the virtual human (e.g., the ability to look and sound like a real cancer patient). Realism in this context is meant to include holistic aspects of the interaction such as lip syncing. Discussions with our medical collaborators indicated the importance of emotional expression in real patient encounters and we therefore incorporated statements to quantify the user's perception of the virtual human's ability to emote. Our dialogue statements were focused on quantifying the quality of the dialogue itself. Aspects of the dialogue such as whether the responses were fluent, natural, relevant, logical and/or emotionally expressive were selected as the criteria. All statements were evaluated on a 1-5 likert scale with 5 representing strong agreement.

Our UI rating scale was also developed by the research team. We simply had the research subjects evaluate the utility of each UI element on the system feedback screen (see Fig. \ref{fig:feedback} using a 1-5 likert scale with 5 representing high utility. Likert scales allows for more nuance \cite{allen2007likert} in survey responses and are appropriate for gathering feedback for system improvement.

\begin{figure} 
    \centering
    \caption{}
    \textbf{Sample excerpt from conversation with SOPHIE}
    \begin{quote}
    \vspace{7.5pt}
    {\fontfamily{pcr} \selectfont 
    
\textbf{SOPHIE}: I've already scanned the results, but don't understand what they are telling
        me. Could you explain what they mean?\\
    \vspace{5pt}
\textbf{user}:   So unfortunately Sophie I have some bad news. It looks like the cancer has
        grown and spread.\\
        \vspace{5pt}
\textbf{SOPHIE}: Those are not the words I wanted to hear. I mean, I was bracing for the worst,
        since I could tell by the pain that it's bad. But to hear that the cancer has
        spread is quite depressing. What does it all mean for me?\\
       \vspace{5pt} 
\textbf{user}:   How much information would you like to know about the prognosis?\\
    \vspace{5pt}
\textbf{SOPHIE}: I feel very anxious about my condition. I feel like it's gotten worse. I want
        to know what this means for me.\\
        \vspace{5pt}
\textbf{user}:   What concerns do you have about the future?\\
    \vspace{5pt}
\textbf{SOPHIE}: What worries me the most is just how uncertain my future is. Some days I
        believe my condition will get better and others I brace myself for the worst.
        Is there any way you can let me know what I should expect?
        
}
\end{quote}
    \label{fig:transcript}
\end{figure}


\section{Results}
\subsection{Ratings Comparison}

\begin{table*}[!htb]
    \caption{Average SP ratings for each item in the rating scale. Italics and an asterisk (``*'') denote $p \leq 0.05$}
    \label{tab:results}
	\centering
	\begin{tabularx}{\textwidth}{c|X|c|c|c}
		\hline
		\textbf{ID} & \textbf{Question} & \textbf{Control Mean} & \textbf{Intervention Mean} & \textbf{p-value} \\
		\hline
		q1 & The participant elicited the patient's major concerns within the first 5 minutes of the conversation. & 0.867 & 1.0 & 0.175 \\
		\hline
		q2 & The participant asked for permission to share information about prognosis. & 0.533 & 0.8 & 0.114 \\
		\hline
		\textit{q3} & \textit{The participant asked how much information the patient would like concerning prognosis.} & \textit{0.333} & \textit{0.8} & \textit{0.03*} \\
		\hline
		q4 & The participant checked the patient's prognostic understanding by asking them to state what they understood, using a teach-back approach. & 0.133 & 0.267 & 0.282 \\
		\hline
		q5 & The participant actively encouraged the patient to ask questions using facilitating questions/statements (e.g., What questions do you have? At this point many patients have questions etc.). & 0.333 & 0.6 & 0.078 \\
		\hline
		q6 & The participant helped the SP make a plan regarding with whom, and when, to convey prognostic information to family members. & 0.2 & 0.467 & 0.123 \\
		\hline
		\textit{q7} & \textit{The participant gave the SP many opportunities to talk.} & \textit{0.467} & \textit{0.867} & \textit{0.024*} \\
		\hline
		\textit{q8} & \textit{Empower Rating} & \textit{5.267} & \textit{6.133} & \textit{0.003*} \\
		\hline
		q9 & The participant described the medical situation (the cancer has spread) clearly and without euphemism or jargon. & 0.8 & 0.8 & 0.488 \\
		\hline
		q10 & The participant shared the prognosis accurately (a few months - less than one year). & 0.6 & 0.667 & 0.476 \\
		\hline
		q11 & The participant used clear language without euphemism or jargon when sharing the prognosis. & 0.533 & 0.667 & 0.252 \\
		\hline
		q12 & The participant used difficult to understand medical jargon. & -0.733 & -0.933 & 0.079 \\
		\hline
		q13 & The participant lectured the patient (uninterrupted information for what seemed like a long time). & -0.533 & -0.8 & 0.067 \\
		\hline
		q14 & be Explicit rating & 5.667 & 6.067 & 0.084 \\
		\hline
		\textit{q15} & \textit{The participant was generally empathetic.} & \textit{0.667} & \textit{1.0} & \textit{0.04*} \\
		\hline
		q16 & The participant used states of empathy. & 0.467 & 0.667 & 0.205 \\
		\hline
		q17 & The participant used silence appropriately in response to patient emotion. & 0.333 & 0.667 & 0.051 \\
		\hline
		\textit{q18} & \textit{The participant validated the SP emotional responses.} & \textit{0.533} & \textit{0.867} & \textit{0.027*} \\
		\hline
		q19 & Empathize Rating & 5.533 & 6.0 & 0.102 \\
		\hline
		\textit{q20} & \textit{Overall Communicator} & \textit{5.067} & \textit{6.0} & \textit{0.003*} \\
		\hline
		-- & \textbf{Total} & \textbf{29.6} & \textbf{36.067} & \textbf{0.005*} \\
		\hline
	\end{tabularx}
\end{table*}

We used four SPs for the experiment.  Each SP had an equal number of intervention and control participants ($\pm 1$). A Bonferroni corrected pairwise t-test showed no significant differences between ratings given by the different SPs. 

We found that the intervention group performed significantly better on the ``overall communicator" (intervention: 6.000, control: 5.067, $p<0.05$) and ``aggregate score" (intervention: 36.067, control: 29.600 $p<0.05$) metrics. For every other question, there was a trend towards the intervention group, but the difference was not always statistically significant. See Table \ref{tab:results} for the full results.

\subsection{UI/UX Surveys}
Participants in the intervention group rated each feedback metric shown in figure \ref{fig:feedback} on a 1-5 Likert scale and the results can be seen in table \ref{tab:UI_feedback}. Overall, we found that the most useful feedback metrics were the reading level, speaking rate, hedge words, transcript and turn-taking. The least useful metrics were positive emotion, empathy words and personal pronouns. It is important to note that user's ratings of the feedback metrics may not equate to what they actually learned. For example, participants rated the empathy metrics relatively low, yet still performed significantly better on 2 of the 5 empathy ratings according to the Human SP evaluation. 

\begin{figure*}[h]
    \centering
    \includegraphics{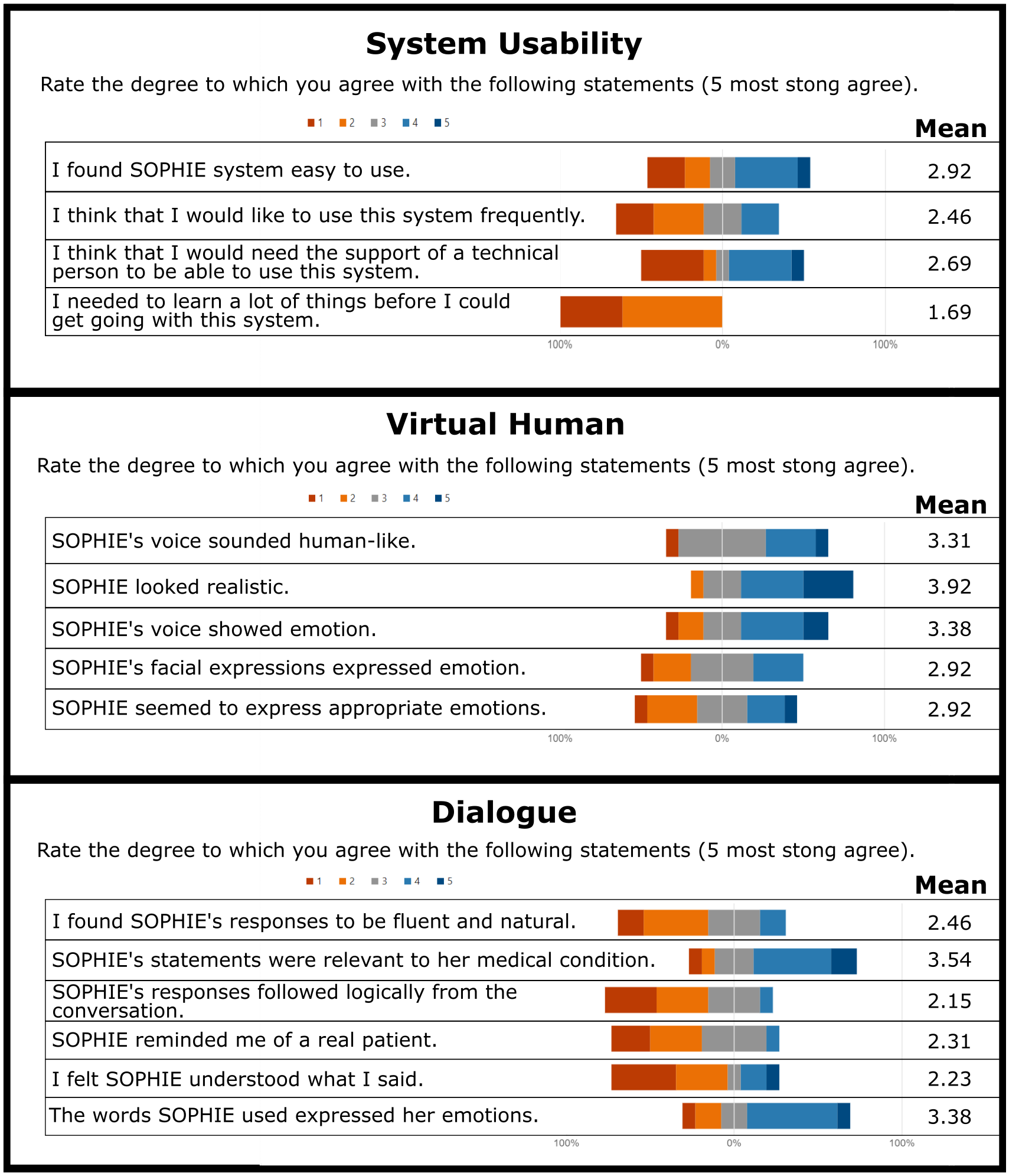}
    \caption{\textbf{UX survey responses - } \textit{System Usability:} The system appears easy to use, however, participants would not use system frequently. \textit{Virtual Human:} SOPHIE looked realistic, is capable of showing emotion through voice, yet lacks ability to express emotions through facial expression. \textit{Dialogue:} SOPHIE's responses were relevant to her medical condition. However, she did not appear to understand the user and her responses were judged as not fluent or natural, illogical, and unlike a real patient.}
    \label{fig:UX_responses}
\end{figure*}

Additionally, participants rated four components of the SOPHIE system; System Usability, Virtual Human, Dialogue, and Feedback. Every question was likewise asked on a 1-5 point Likert scale with 5 meaning most strongly agree. Fig. \ref{fig:UX_responses} depicts the UX experience for system usability, virtual human and dialogue. Overall, participants rated the system as easy to use with the virtual human having a realistic voice and appearance. However, the dialogue appears to be a major point of weakness in the experiment. Its responses were not rated as natural, logical, or realistic, and it did not appear to understand the user. Importantly, though, SOPHIE kept the conversation relevant, despite the variety of ways in which users could respond. SOPHIE's ability to display emotion received mixed ratings.

Despite these limitations, debriefing interviews with participants in the intervention group indicated that participants saw overall utility in our system, and participants expressed an interest in using the tool if improvements to the dialogue and virtual human allowed the interaction to be more realistic.

\section{Discussion}

\subsection{Improving Communication Skills}
Human communication is pragmatic, with patterns developing over time to become habitual and difficult to change \cite{watzlawick2011pragmatics}.  We speculate that the extent to which an individual's communication behaviors can be modified is dependent on how well a person's subjective experience and recollection can be meaningfully connected to clearly-presented and actionable feedback. Based on our experiment, we see that the combination of interacting with SOPHIE and receiving automated feedback improved participant's use of the Empower and Empathize skills and their overall communication. To what extent this increase was the result of simply interacting with SOPHIE versus receiving feedback on the interaction cannot be established based on this experiment, as we didn't have a population which had the conversation but not the feedback. However, one indicator of the importance of feedback may be that the users rated the feedback system more highly than the dialogue and virtual human. We suspect that the feedback system is a major contributor to the differences observed. For example, a user, upon reviewing the transcript of their most recent conversation with SOPHIE and observing that they only asked three questions, may realize that they can better empower the patient by asking more questions. Similarly, as the user is reviewing the transcript, they become aware of empathetic statements they could have used. This knowledge seems to inform subsequent conversations based on the ratings from the human SPs (see q3 - asking questions - and q15 - empathetic statements - in table \ref{tab:results}). This suggests that the system's feedback could result in an actionable plan for improvement. Further experiments with SOPHIE will be needed to confirm these intuitions about the efficacy of the feedback system. 

Although we did see differences in the being Explicit skill, they were not statistically significant. This may change as we run future experiments with larger sample sizes.  Additionally, we are planning further improvements to the SOPHIE system based on our UI/UX feedback from study participants (see section \ref{future}). Ultimately, we believe the consistency of the virtual human, dialogue and feedback system would allow a healthcare professional to hone a variety of communication skills through repeated practice.



\subsection{Promoting Equity and Access}
The SOPHIE  communication resources can be made accessible to anybody with a computer, microphone and internet connection. The accessibility of the system is highlighted by the fact that users rated SOPHIE as easy to use, and in particular disagreed with the statement ``I needed to learn a lot of things before I could get going with this system." The scalable, web-based nature of SOPHIE makes it a low-cost alternative to synchronous training with human SPs, and could make communication training more readily available to rural or low-income regions. This would promote equity by making communication training available to more healthcare professionals regardless of the financial resources available to them. 

Specific aspects of SOPHIE could also be customized to reflect a diverse range of patients. Attributes like SOPHIE's age, race, gender, language, and personality could be modified to represent all demographics of patient populations. Additionally, the context of the module could be readily changed. In the future, users could choose from dozens of healthcare modules focusing on specific types of conversations with customized virtual humans uniquely suited for the purpose. Different types of patient personalities could be programmed to help practice responding to different reactions and attitudes from diverse patients.

\begin{table}[h!]
     \caption{\textbf{User ratings for feedback metrics - } All metrics were rated using a 1-5 Likert scale for the statement ``I found the $<$metric$>$ feedback useful." The most useful metrics were reading level, speaking rate, and hedge words. The least useful metrics were positive emotion, empathy words and personal pronouns.}
    \centering
    
    \begin{tabularx}{\columnwidth}{X|c}
        \hline
        \textbf{Feedback Component} & \textbf{Average rating} \\
        \hline
        Questions Asked & 3.385 \\
        Open Ended Questions & 3.538 \\
        Turn taking & 3.846 \\
        \textit{``Empower" metrics total} & \textit{3.590} \\
        \hline
        Personal Pronouns & 3.231 \\
        Empathy Words & 3.077 \\
        Positive Emotion & 2.769 \\
        \textit{``Empathize" metrics total} & \textit{3.026} \\
        \hline
        Hedge Words & 3.846 \\
        Speaking Rate & 4.000 \\
        Reading Level & 4.231 \\
        \textit{``be Explicit" metrics total} & \textit{4.026} \\
        \hline
        Transcript & 3.846 \\
        Suggestions in Transcript & 3.538 \\
        \hline
    \end{tabularx}
    \label{tab:UI_feedback}
\end{table}

\subsection{Future of SOPHIE}
\label{future}
Despite efforts to improve communication skills, a clinician may fall back into their old habits unconsciously. Thus, there is a need to consistently practice these difficult conversations for maintaining, or, even enhancing, a medical professional's skill proficiency. Future generations of SOPHIE aim to satisfy this need by improving upon the system weaknesses discovered from the UI/UX responses: namely, the dialogue management as well as the emotional expressiveness of the SOPHIE virtual human. Additionally, we plan to iterate further with medical professionals to perfect the feedback system (especially in regards to feedback elements that received low scores). 

The feedback system may be extended to other applications in the future as well. For example, clinicians could have an application on their phones that could be used during real patient encounters. With the patient's consent, a clinician could use the app to analyze the conversation. The app could generate a checklist that the clinician could quickly review to determine if they need to spend more time addressing a specific area with their patient. The data could even be tracked over time to help the clinician monitor their performance and obtain user-led, personalized insights. For example, the clinician could view the system's recommended SOPHIE modules to refresh or improve certain communication skills. What the clinician decides to do based on this feedback is entirely their own and the design focus should always be to empower the clinician.

\subsection{Limitations}
The limitations for this study are our small sample size, lack of clarity of which factors of the SOPHIE system caused the improvements and a lack of formal validation for the rating scales used. We elaborate on these limitations in our ethical statement which can be found after the conclusion. 

\subsection{Contributing to SOPHIE}
Regretfully, the SOPHIE code base is not open-source. The SOPHIE project is an ongoing venture between the University of Rochester Computer Science (URCS) Department and the University of Rochester Medical Center (URMC). Once deployed, the research staff may release a starter kit for researchers who wish to run similar experiments. However, we are actively recruiting participants for clinical trials and seeking additional collaborations from other medical schools. If interested, please email Kurtis Haut at \textit{khaut@u.rochestetr.edu}.


\section{Conclusion}
We developed and validated a new digital tool for improving serious illness communication training for health care professionals. We observed significantly better performance on overall communication and higher aggregate scores as a result of interacting with SOPHIE. This study suggests the potential for practicing conversations with  a virtual human and an automated feedback system to improve communication skills in a scalable, on-demand fashion. By improving access to communication training, SOPHIE  could improve the equity of our local institutions, and perhaps even the global healthcare system.

\section{Ethical Impact Statement}
The impacts of this system will be felt by real patients, whose experiences of receiving tragic news will be shaped by the behaviors their clinicians developed during their communication training. Given the effect that communication has on patient healthcare outcomes (see section \ref{intro}), the ethical considerations of this work must be taken seriously. We have an obligation to ensure the efficacy of the SOPHIE system and mitigate any potential harms it could cause. Thus, the virtual patient dialogue and the automated feedback must be based on the highest standard of established medical practices. Otherwise, learned communication deficiencies could cause patient prognosis misunderstandings, leading to healthcare choices that are unaligned with patient values. To mitigate this risk, the development of SOPHIE has been heavily shaped through several design iterations with expert oncologists, palliative care specialists and other stakeholders. We plan to conduct further experiments to validate and continually refine each subsequent generation of SOPHIE before deploying the final version. We believe that continuous system evaluations that appropriately keep pace with development will help maintain high ethical standards.  

Careful ethical considerations must also be made in experimental design for the validation process and ensure participant confidentiality. In our experiment, all participants provided informed, written consent before beginning the study. The methods were performed in accordance with relevant guidelines and regulations that were approved by our university's Institutional Review Board. All data collected has been de-identified such that it can not be traced to a specific participant. 
 
One limitation of our study is the small, relatively homogeneous sample size consisting of predominantly white healthcare professionals from our local area which reduces the generalizability of our results.  Future experiments will aim to recruit a larger, more demographically diverse sample to ensure that the needs of all users are met. 

Additionally, the rendering of the virtual human is a potential source of bias. We are depicting a white, elderly, female with terminal lung cancer and set personality. This could pose an ethical issue because real patients are demographically diverse and come from a variety of backgrounds. It may be the case that communication skills learned from interacting with SOPHIE do not translate perfectly when communicating with patients who do not match SOPHIE's race, gender, age or personality. Making these features more customizeable will be a focus of future iterations of the system. This would help healthcare professionals prepare to communicate equally well regardless of their patient's demographic traits, background or personality.

\section{Data Availability}
De-identified results from the experiment are available upon request. 


\bibliographystyle{IEEEtran}
\bibliography{references}



\end{document}